\documentstyle[multicol,prl,aps,psfig,floats]{revtex}
\begin{document}
\tightenlines
\title{Comparison between Theory and Some Recent Experiments on Quantum 
Dephasing}
\author{Dmitrii S. Golubev and Andrei D. Zaikin}
\address{Forschungszentrum Karlsruhe, Institut f\"ur Nanotechnologie,
76021 Karlsruhe, Germany\\
I.E.Tamm Department of Theoretical Physics, P.N.Lebedev
Physics Institute, Leninskii pr. 53, 117924 Moscow, Russia}

\maketitle

\begin{abstract}
We report on a quantitative comparison between our theory of quantum
dephasing at low temperatures and some recent experimental results
[D. Natelson {\it et al.}, Phys. Rev. Lett. {\bf 86}, 1821
(2001); A.B. Gougam {\it et al.}, J. Low Temp. Phys. {\bf 118}, 447 (2000);
F. Pierre {\it et al.} cond-mat/0012038].

\end{abstract}


\begin{multicols}{2}


Experiments \cite{Webb}
demonstrated low temperature saturation of the electron decoherence
time $\tau_{\varphi}$ in mesoscopic conductors. This result attracted a lot of
attention and triggered theoretical debates on a fundamental issue
of quantum dephasing at zero temperature. More recently new experiments in
various mesoscopic systems were performed by several groups. The results
of these studies are still waiting for a careful analysis.

In this note we will restrict our attention to 
the results reported by two groups \cite{Nat} and
\cite{SM1,SM2}. Natelson {\it et al.} \cite{Nat}
found that their data for $\tau_{\varphi}$ are inconsistent
with the geometry dependence predicted in a standard theory
\cite{AAK}. For wider samples they observed saturation 
of $\tau_{\varphi}$ at low temperatures. The Saclay-Michigan
collaboration detected no saturation in silver \cite{SM1} and
pure gold \cite{SM2} samples down to $45\div 50$ mK, but 
observed clear saturation in copper at temperatures one order
of magnitude higher. It is the purpose of this note
to analyze if the results \cite{Nat,SM1,SM2} can be explained
within our theory of quantum dephasing \cite{GZ98}.

In order to do so let us briefly recollect our key results. The 
electron decoherence time $\tau_{\varphi}$ in diffusive conductors was 
predicted \cite{GZ98} to be finite at any temperature including $T=0$ due to 
electron-electron interactions. In simple terms our result can be 
expressed as follows
\begin{equation}
\frac{1}{\tau_{\varphi}(T)}=\frac{1}{\tau_{\varphi 0}}+\frac{1}{\tau (T)},
\label{sum}
\end{equation}
where $\tau_{\varphi 0}$ is the electron dephasing time at zero temperature
and $\tau (T)$ is basically the dephasing time calculated in \cite{AAK}
(note, however, that  $\tau (T)$ depends on the full dephasing time
$\tau_{\varphi}(T)$, so the relation (\ref{sum}) should be understood as the
equation for $\tau_{\varphi}(T)$). The value $\tau_{\varphi 0}$ is 
predominantly determined by {\it high-frequency}
modes of the electromagnetic field which mediates interaction between
electrons in disordered conductors. One has \cite{GZ98}
\begin{equation}
\frac{1}{\tau_{\varphi 0}} \propto \omega_c^{d/2},
\label{cutoff}
\end{equation}
where $\omega_c$ is the high-frequency cutoff and $d$ is the effective
dimension. It is important to emphasize that $\omega_c$ {\it cannot} exceed
$1/\tau_e$, where $\tau_e=l/v_F$ is the electron elastic scattering time.
This is because at frequencies above $1/\tau_e$ the electron motion
is ballistic and the system is noiseless at such scales. 
Therefore no contribution from $\omega >1/\tau_e$ to the noise correlator 
(which determines $\tau_{\varphi}$) can occur. On the other hand, for
$\omega  \ll 1/\tau_e$ the electron motion is diffusive and the noise
correlator is finite decaying to zero as $\omega$ approaches $1/\tau_e$.
Thus we choose \cite{FN}
\begin{equation}
\omega_c=a_d/\tau_e,
\label{cutoff2}
\end{equation}
where $a_d$ is some numerical factor $a_d \leq 1$. In \cite{GZ98} we have set
$a_d=1$. This was sufficient for the order-of-magnitude estimate for
$\tau_{\varphi 0}$. Note, however, that since this sharp cutoff procedure 
slightly {\it overestimates} the contribution of
frequencies close to $\omega_c$, the
actual value of $\tau_{\varphi 0}$ is expected to be bigger than that
of Ref. \cite{GZ98} by a numerical factor of order one. 

It is also worth mentioning that, provided the elastic mean free 
path $l$ is smaller
than the thickness $t$ and the width $w$ of the sample, 
$l \ll w,t$, our $3d$ expression for the dephasing rate
\begin{equation}
\frac{1}{\tau_{\varphi 0}^{(3d)}}= \frac{e^2\rho}{3\pi^2 \sqrt{2D}} \left(
\frac{a_3}{\tau_e}\right)^{3/2} \label{3d}
\end{equation}
should be used even if the sample is quasi-$1d$ (quasi-$2d$) in a standard
sense, i.e. if both $L_{\varphi }=\sqrt{D\tau_{\varphi}}$ and   
$L_{T}=\sqrt{D/T}$ exceed $t$ and $w$ (or $t$). Here
$\rho$ and $D$ are respectively the resistivity and the diffusion coefficient.
Only if 
$l \gtrsim w,t$ ($l \gtrsim t$) our 1d (2d) expressions should be applied.
The reason for that was discussed after eq. (81) of the second Ref. \cite{GZ98}.
In short, eq. (\ref{3d}) is relevant since 
$\tau_{\varphi 0}$ is dominated by high frequencies at which the electron 
diffusion is always 3-dimensional provided  $l \ll w,t$. Of course, eq. 
(\ref{3d}) crosses over to $1d$ (or $2d$) results for  $l \sim w \sim t$ 
(or $l \sim t$), as is seen from the following relations (presented at $a_d=1$)
\begin{equation}
\frac{\tau_{\varphi 0}^{(1d)}}{\tau_{\varphi 0}^{(3d)}}=\frac{tw}{2\pi
l^2},\;\;\;\;\;\;
 \frac{\tau_{\varphi 0}^{(2d)}}{\tau_{\varphi 0}^{(3d)}}=\frac{4}{\pi
\sqrt{6}}\frac{t}{l}.
\label{1d2d}
\end{equation}

We are now prepared to compare experimental results
\cite{Nat,SM1,SM2} with our theoretical predictions. We will address
both the maximum values of the decoherence time 
$\tau_{\varphi}^{max}$ reached in these
experiments at the lowest measurement temperatures and the observed 
temperature dependencies of the dephasing time depending on the sample
parameters.

\section{Maximum dephasing times}

For all the samples reported in Refs. \cite{Nat,SM1,SM2} the mean free path $l$
deduced from the values $D$ was 2 to 5 times {\it
shorter} than both $t$ and $w$ for quasi-$1d$ samples and $t$  for
four quasi-$2d$ samples of Ref. \cite{Nat}. Therefore, even though the
condition $l \ll w,t$ was not very well satisfied in these experiments, it
would be more appropriate  to use eq.
(\ref{3d}) for our comparison. It is also easy to observe from eqs.
(\ref{1d2d}) that, since $l$ is not drastically smaller than $w$ and $t$,  
our $1d$ and $2d$ expressions will yield the same order of magnitude estimates
for $\tau_{\varphi}^{max}$ for all the samples in question (also see below). 

The maximum dephasing times were reported in Ref. \cite{Nat} for ten
(six quasi-$1d$ and four quasi-$2d$) samples fabricated from the same material
($AuPd$) with practically the same resistivity. For all ten samples the maximum
dephasing times were found to be nearly universal $\tau_{\varphi}^{max} \approx
(0.8 \div 2)\times 10^{-11}$ sek {\it independently} of the sample geometry.
At $T \sim 80$ mK these values yield $T\tau_{\varphi} \approx 0.09\div 0.22$.
Since both the material and its properties were the same an important test
for our theory is to fit the data \cite{Nat} for $\tau_{\varphi}^{max}$
to eq. (\ref{3d}) which only depends on the material parameters and not on
geometry.

In experiments \cite{SM1,SM2} the maximum dephasing times were found to be 
2 to 3 orders of magnitude {\it bigger} as compared to those in Ref.
\cite{Nat}. Furthermore, $\tau_{\varphi}^{max}$ for the silver sample
\cite{SM1} was about an order of magnitude higher than that for the copper
sample \cite{SM1} with similar (although {\it not} identical) parameters.
Observing this difference the authors \cite{SM1} suggested that the low
temperature saturation of $\tau_{\varphi}$ in disordered metal wires cannot be
universal and can be material dependent.

For our comparison we will use the data for 8 samples \cite{Nat} with nominally
identical resistivity $\rho \approx 24$ $\mu \Omega$ cm and $D 
\approx 1.5 \times 10^{-3}$ m$^2$/sek (samples
C to F and H to K), 2 samples \cite{SM1} ($Ag$ and $Cu$) and one sample
($AuMSU$) \cite{SM2}. Since all the maximum dephasing times \cite{Nat} were
nearly the same, for the sake of brevity we only quote an average value
for  $\tau_{\varphi}^{max}$ obtained by averaging over samples C to F and H to
K.  In order to estimate $\tau_{\varphi 0}$ for $Ag$ and $Cu$
samples we used the parameters given in the Table 1 of Ref. \cite{SM1}.
Experimentally observed values of $\tau_{\varphi}^{max}$ and our
theoretical predictions for $\tau_{\varphi 0}$ (eq. (\ref{3d}) with 
$a_3=1$) are summarized in the following Table:

\begin{center}
\begin{tabular}{|c|c|c|c|c|}
\hline 
sample & $\tau_{\varphi}^{max}$ (sek)& $\tau_{\varphi 0}$ (sek)\\
\hline   C to F and H to K \cite{Nat} (averaged)&  1.3$\times 10^{-11}$  & 
0.3$\times10^{-11}$\\ \hline   $Ag$ \cite{SM1} & 10$\times 10^{-9}$  & 
2$\times 10^{-9}$   \\ \hline   $Cu$\cite{SM1}  & 1.8$\times 10^{-9}$  &  
0.3$\times 10^{-9}$\\ \hline   
$AuMSU$ \cite{SM2}  & 8$\times 10^{-9}$  & 4$\times 10^{-9}$\\ \hline  
\end{tabular} 
\end{center}

These results demonstrate that for {\it all} the above samples our formula
(\ref{3d}) with $a_3=1$ gives correct order-of-magnitude estimates for  
$\tau_{\varphi}^{max}$ even though these times differ by 2 to 3 orders of
magnitude depending on the sample and experiment. Already this
agreement is fairly good since in weak localization
measurements $\tau_{\varphi}$ is {\it defined} up to a numerical
prefactor. Furthermore, in accordance with our
expectations, in all cases our theoretical estimates involving the cutoff
parameter $a_3=1$ are {\it smaller} than the experimental values 
by a numerical factor ranging from 2 to 6. This implies that
by choosing the cutoff parameter in the range $a_3 \approx 0.3 \div 0.6$
one can exactly reproduce all the above experimental values for 
$\tau_{\varphi}^{max}$ from {\it the same} formula (\ref{3d}) which only depends
on the material parameters, like $\rho$ and $D$.

One can also eliminate the ambiguity related to the cutoff parameter $a_d$
in (\ref{3d}) by calculating the ratio between maximum dephasing times for
different samples. Making use of the experimental results \cite{Nat,SM1,SM2}
and our theoretical estimates based on eq. (\ref{3d}) one finds:

\begin{center}
\begin{tabular}{|c|c|c|c|c|}
\hline 
 & experiment & theory (eq. (\ref{3d}))\\ \hline 
$\tau_{\varphi 0}^{AuPd}/\tau_{\varphi 0}^{Ag}$    & $1.3\times
10^{-3}$   & $1.5\times 10^{-3}$\\ \hline  
$\tau_{\varphi 0}^{AuPd}/\tau_{\varphi 0}^{Cu}$    & $0.7\times
10^{-2}$   & $1\times 10^{-2}$\\ \hline
$\tau_{\varphi 0}^{AuPd}/\tau_{\varphi 0}^{AuMSU}$    & $1.6\times
10^{-3}$   & $0.8\times 10^{-3}$\\ \hline
\end{tabular} 
\end{center}

The agreement appears to be even much better than one could 
possibly expect given all experimental and theoretical uncertainties in
determination of $\tau_{\varphi}$. We emphasize that our
theoretical predictions presented in both above Tables involve
{\bf no fit parameters}. Previously \cite{GZ98} we already reported on a
similar agreement with the results \cite{Webb}. All these observations 
strongly favour the conclusion about {\it universality}  of the low
temperature saturation of $\tau_{\varphi}$ in disordered conductors.

To complete this issue we mention that also our $1d$ and $2d$ expressions
for the  maximum dephasing time \cite{GZ98} give correct
order-of-magnitude estimates for $\tau_{\varphi}$ for the above samples. For
instance, for $2d$ samples \cite{Nat} the second eq. (\ref{1d2d}) yields (for
$a_2=1$) $\tau_{\varphi 0}^{(2d)} \sim 0.6\times 10^{-11}$ sek. This estimate
is only $2\div3$ times smaller than the experimental values. Again, although
such an accuracy is already more than sufficient, the remaining 
difference can easily be removed simply by choosing $a_2 \approx 0.3 \div 0.5$.

For comparison, the
authors \cite{Nat} estimated the value of $\tau_{\varphi}$ which follows
from the theory \cite{AAK} for their $2d$ samples as 
$\tau_{\varphi} \sim 6 \times 10^{-9}$ sek at $T=1K$. At the lowest
measurement temperature $T \approx 80$ mK this estimate translates into
$\tau_{\varphi}$ in the range $10^{-7}$ sek which is $3.5\div 4$ orders
of magnitude bigger than the values measured in Ref. \cite{Nat}.

\section{Temperature Dependence}

We now analyze the temperature dependence of $\tau_{\varphi}$ detected
in experiments \cite{Nat,SM1,SM2}. For quasi-$1d$ samples 
eq. (\ref{sum}) takes the form \cite{GZ98,FN1}
\begin{equation}
\frac{1}{\tau_{\varphi}}=\frac{1}{\tau_{\varphi 0}}
+\frac{\rho T}{R_q tw}\sqrt{2D\tau_{\varphi}},
\label{eq1d}
\end{equation}

\end{multicols}
\begin{figure}
\centerline{\psfig{file=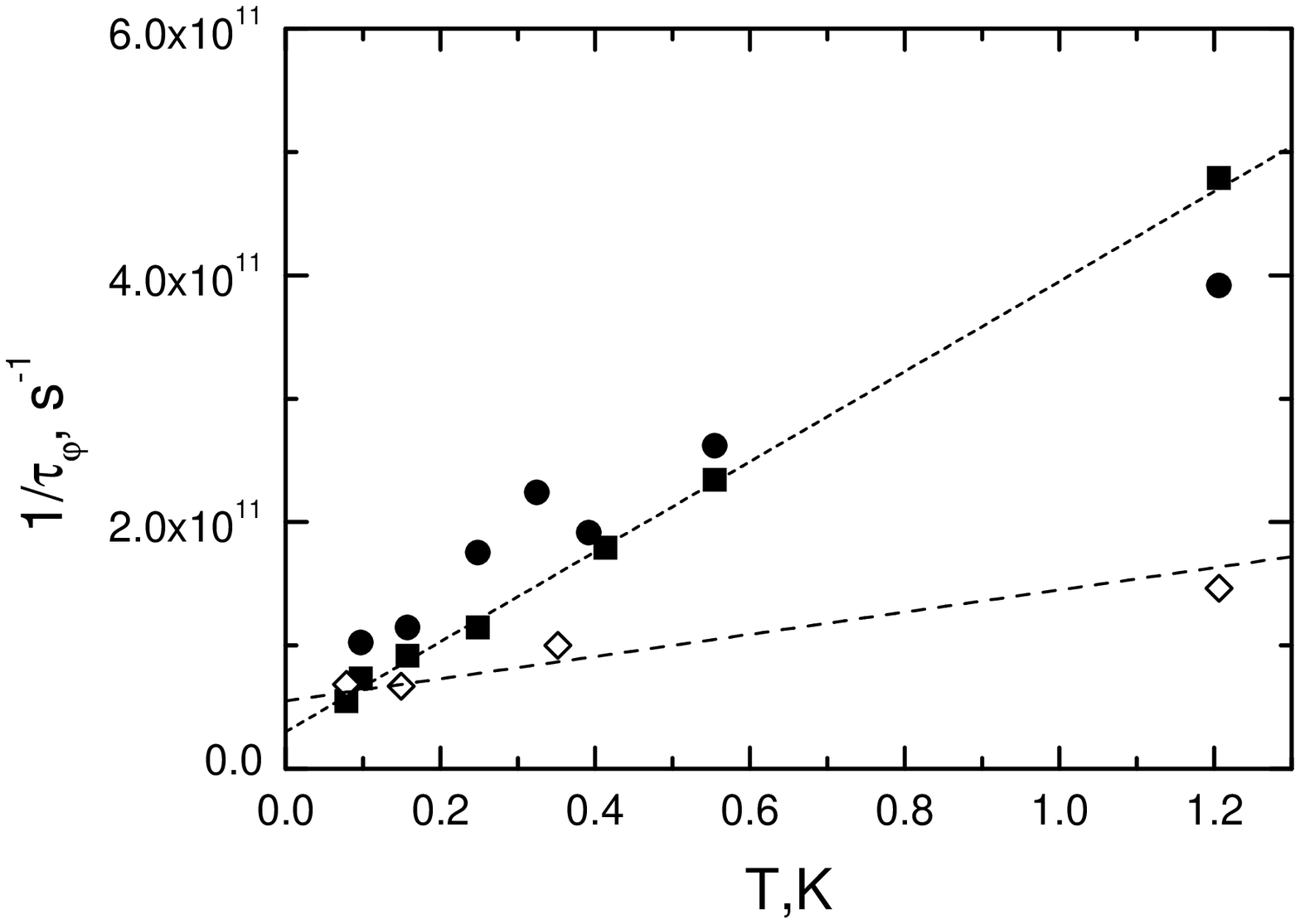,width=12cm}}
\caption{Dephasing rate $1/\tau_{\varphi}$ measured in Ref. [2] for the
samples A (squares), B (circles) and F (open symbols) as a function of
temperature $T$.} 
\label{fig1}
\end{figure}

\begin{multicols}{2}

\noindent
where $R_q=\pi/2e^2 \simeq 6453$ $\Omega$.
An essential difference between this result and that of Ref. \cite{AAK} is
that in our case the temperature-independent contribution to the
dephasing rate $1/\tau_{\varphi 0}$ is not equal to zero. If one neglects this
contribution, eq. (\ref{eq1d}) would immediately yield the standard
result \cite{AAK} $1/\tau_{\varphi AAK}^{(1d)} \propto T^{2/3}$. Similarly, 
in quasi-$2d$ samples one gets \cite{AAK} 
$1/\tau_{\varphi AAK}^{(2d)} \propto T\ln T$. In order to test these
predictions it is appropriate to present experimental data for
$\tau_{\varphi}(T)$ on a log-log plot, exactly as it was done in Refs.
\cite{Nat,SM1,SM2}. From Ref. \cite{AAK} one expects ($i$) to observe straight
lines with the slope becoming {\it steeper} as one crosses over from
quasi-$1d$ to quasi-$2d$ systems and ($ii$) for two quasi-$1d$ samples with
the same $\rho$ and $D$ but with different cross-sections $tw$ the straight
lines corresponding to $T^{-2/3}$-dependence of $\tau_{\varphi}$ should be {\it
parallel}.

None of these two features was observed in the experiments \cite{Nat}.
Just on the contrary, the opposite trend was detected: In all cases the
slope of the curves {\it decreased} with increasing sample width $w$ (see Fig.
3 of Ref. \cite{Nat}). The curves for quasi-$1d$ samples with different $w$
were {\it not} parallel (actually the data points for the sample F on Fig. 3
of \cite{Nat} clearly saturate and can hardly be fitted to a straight line at
all) and, on top of that, the $T$-dependence of the decoherence rate for
quasi-$2d$ samples -- instead of being linear in $T$ -- was very weak
already at $T \sim 1K$. 

Now let us see if these features can be accounted for within our theory.
Since the low temperature behavior of $\tau_{\varphi}$ predicted in Ref.
\cite{GZ98} is different from a power-law one \cite{AAK}
it appears to be more useful to replot the same data on a linear scale.
Then, according to eq. (\ref{eq1d}) at sufficiently low $T$
one expects to observe a linear dependence of the decoherence rate on $T$
shifted upwards by the value $1/\tau_{\varphi 0}$. For the same parameters
$\rho$ and $D$ (and the same $\tau_{\varphi 0}$) the slope 
$d(1/\tau_{\varphi})/dT$
for quasi-$1d$ samples should depend only on the cross section $tw$. In
particular, since $w$ for the sample $F$ was reported in \cite{Nat} to be 4
times larger than for the samples A to E, one expects  $d(1/\tau_{\varphi})/dT$
to be 4 times {\it smaller} for the sample F.

The experimental data for the samples A, B and F \cite{Nat} are presented in
Fig. 1. We observe that ($a$) the dephasing rate $1/\tau_{\varphi}$ increases
linearly  with $T$, ($b$) at $T \to 0$ these linear dependencies extrapolate
to a {\it nonzero} (and practically the same) value $1/\tau_{\varphi}$ for all
three samples and ($c$) the slope $d(1/\tau_{\varphi})/dT$ is indeed about 4
times smaller for the sample $F$ as compared to the samples A and B. All these
features are fully consistent with eq. (\ref{eq1d}). The magnitude of the
slope is several ($\sim 4\div 6$) times larger than predicted by eq.
(\ref{eq1d}). The origin of this difference is not clear to us at the
moment, but possibly it can be due to non-uniformity of the wires. The data for
quasi-$2d$ samples I and K \cite{Nat} also show the trend
qualitatively consistent with our eq. (\ref{sum}). Also in that case the
temperature effect seems to be more pronounced than it is predicted within
our theory. 

We conclude that our theory \cite{GZ98} accounts for all the main
experimental observations of Ref. \cite{Nat}.

Now we will address the results of Refs. \cite{SM1,SM2}. In
Fig 3. of \cite{SM1} the authors presented their data for $Au$, $Ag$ and $Cu$
samples with similar parameters. Both the magnitude and the temperature 
dependence of $\tau_{\varphi}$ differ drastically for these three
samples. This observation led the authors \cite{SM1} to conclude
that the behavior of $\tau_{\varphi}$ and, in particular, its saturation
at low $T$ may be material dependent. 

\end{multicols}
\begin{figure}
\centerline{\psfig{file=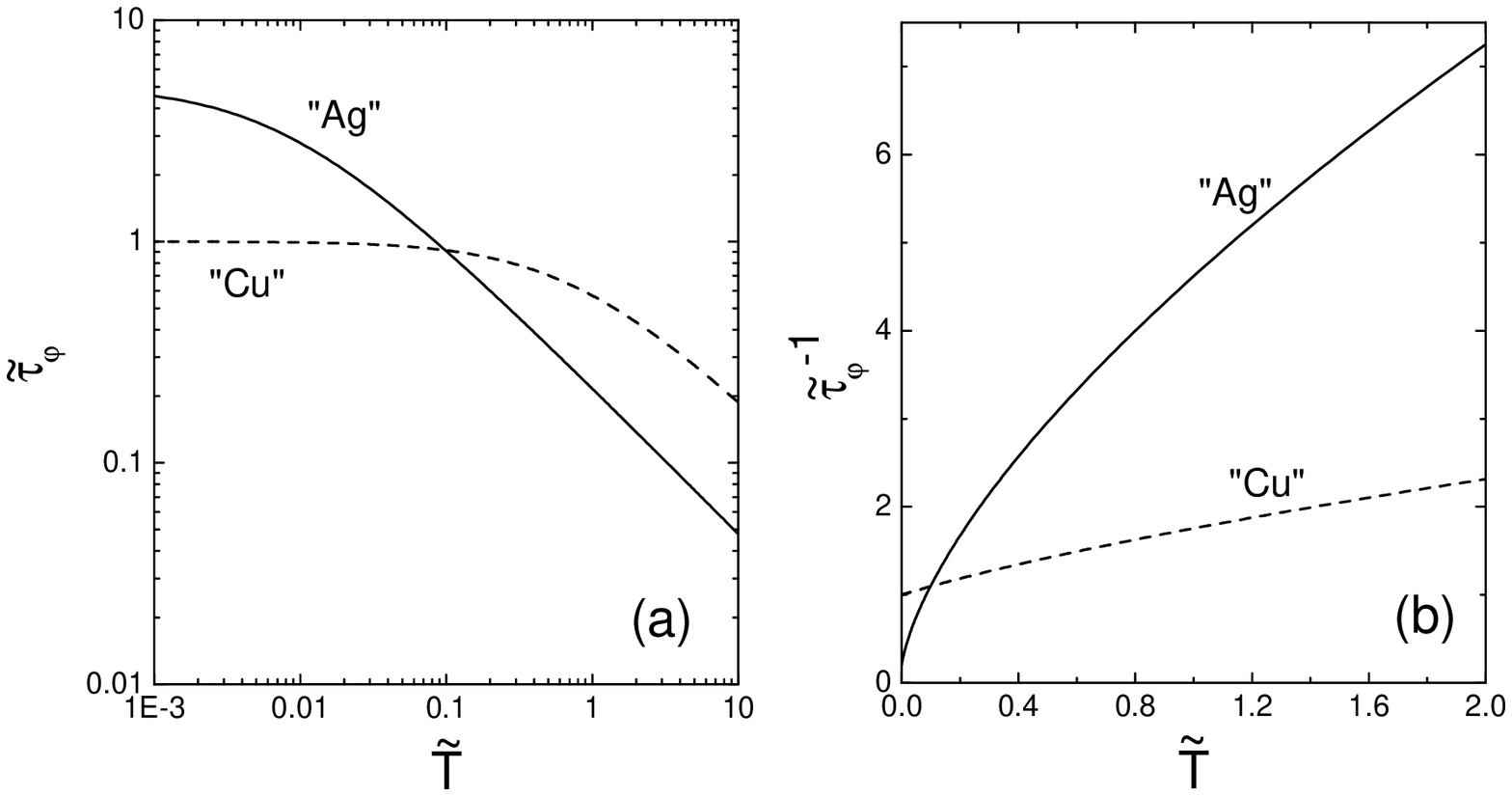,width=18cm}}
\caption{Eq. (\ref{besr2}) plotted on log-log (a) and linear
(b) scales for the parameters corresponding to $Cu$ and $Ag$ samples
of Ref. [3]. For both curves $\tau_{\varphi}$ and $T$ are normalized
respectively to $\tau_{\varphi 0}^{Cu}$ and $T_0^{Cu}$.}  
\label{fig2}
\end{figure}

\begin{multicols}{2}

Later it was demonstrated by the same
group \cite{SM2} that unusually low values of 
$\tau_{\varphi}$ in gold \cite{SM1} were most likely due to high
concentration of magnetic impurities. A pure gold sample ($AuMSU$) was
fabricated and similar behavior of $\tau_{\varphi}$ was
observed \cite{SM2} as previously for $Ag$ sample \cite{SM1}: 
$\tau_{\varphi}$ did not saturate down to $T \sim 45\div 50$ mK
on a log-log plot.  At the same time it was confirmed  
\cite{SM2} that also very pure $Cu$ samples showed saturation
similarly to an earlier $Cu$ sample \cite{SM1}. 

The material dependence of $\tau_{\varphi}$ cannot be ruled out in general.
However, with minimum efforts one can demonstrate that {\it no material
dependence} needs to be assumed in order to quantitatively explain seemingly
different behavior of $Ag$ (no saturation down to $T \sim 49$ mK) and $Cu$
(clear saturation already at $T \gtrsim 700$ mK) samples \cite{SM1}. 

Let us rescale eq. (\ref{eq1d}) as
\begin{equation}
1/\tilde \tau_{\varphi}=A+B\tilde T\sqrt{\tilde \tau_{\varphi}},
\label{besr2}
\end{equation}
where we introduced dimensionless variables
\begin{equation}
\tilde \tau_{\varphi}=\frac{\tau_{\varphi}}{\tau_{\varphi 0}},
\;\;\;\;\; \tilde T=\frac{T}{T_0},\;\;\;\;\;T_0=\frac{R_qtw}{\rho
\tau_{\varphi 0}^{3/2} \sqrt{2D}}.
\label{besr}
\end{equation}
Here $A=B=1$ both for $Cu$ and $Ag$ samples provided one performs
the scaling to the values $\tau_{\varphi 0}$ and $T_0$ calculated for 
the respective sample. The temperature $T_0$ sets the scale at which thermal
and quantum contributions to the dephasing rate become comparable.
Making use of the parameter values for $Cu$ and $Ag$ samples one finds 
\begin{equation}
T_0^{Cu}/T_0^{Ag} \sim 20.
\label{Ts}
\end{equation}
Note, that this estimate is not sensitive to various 
ambiguities related, e.g., to the cutoff parameter $a_d$. Eq.
(\ref{Ts}) 
just follows from eqs. (\ref{3d}), (\ref{eq1d}) for the values
of $\rho$, $D$, $t$ and $w$ presented in the Table 1 of Ref. \cite{SM1}.
The estimate (\ref{Ts}) remains unchanged if, instead of our eq. (\ref{3d}),
one uses experimental values for $\tau_{\varphi}^{max}$ \cite{SM1}. Hence, 
the temperature range where quantum effects set in
can easily vary by one order of magnitude or more for different samples even
if their macroscopic parameters are similar \cite{FN2}.

What remains is to make both log-log and linear scale plots of eq.
(\ref{besr2}), see Fig. 2. In these plots we scaled both curves to the
parameters of one ($Cu$) sample, i.e. $A_{cu}=B_{Cu}=1$, $A_{Ag}=\tau_{\varphi
0}^{Cu}/\tau_{\varphi 0}^{Ag}\simeq 0.2$ and
$B_{Ag}=A_{Ag}^{1/2}T_0^{Cu}/T_0^{Ag} \simeq 9.5$. Making use of the values
for $\tau_{\varphi}^{max}$ \cite{SM1} we find $T_0^{Cu} \approx 730$ mK and
$T_0^{Ag} \approx 40$ mK. These numbers as well as the temperature dependence
of $\tau_{\varphi}$ are in the agreement with experimental findings \cite{SM1}.
E.g. on the log-log plot we observe that, while $Cu$ sample shows clear
saturation around $\tilde T \sim 1$ (i.e. $T \sim 700$ mK), practically
no signs of saturation are visible for $Ag$ sample even at temperatures
one order of magnitude lower. Similar analysis can be performed, e.g., for
$Au6$ \cite{Webb} and $AuMSU$ \cite{SM2} samples (cf. right panel of Fig. 7 in
Ref. \cite{SM2}). In that case one can estimate $T_0^{Au6}/T_0^{AuMSU}
\approx 6$.

\end{multicols}

\begin{figure}
\centerline{\psfig{file=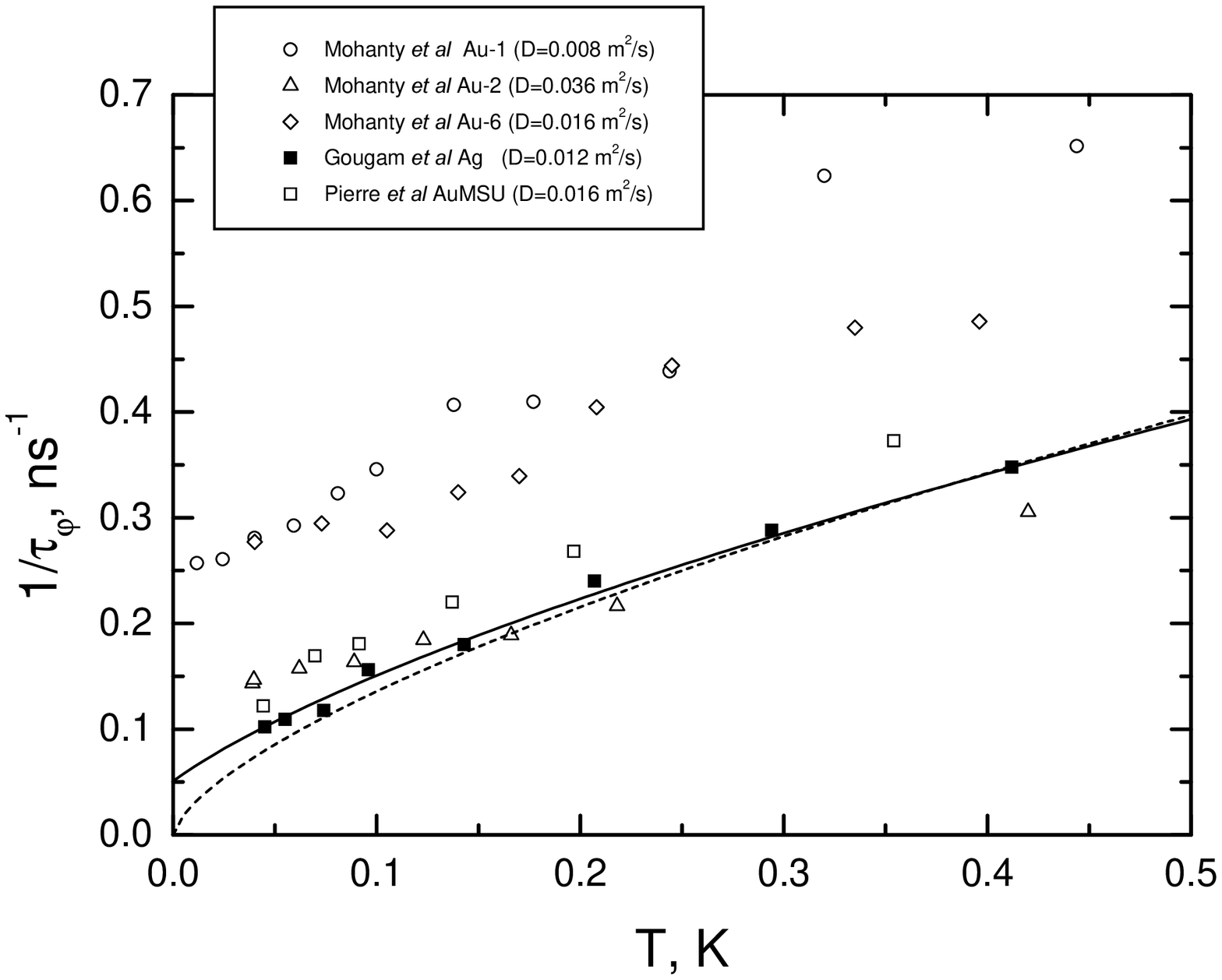,width=12cm}}
\caption{The data for $\tau_{\varphi}(T)$ for several samples from Refs.
[1,3,4]. Dashed and solid lines are the best fits for $Ag$ sample
[3] respectively to theories [5] and [6].} \label{fig3} \end{figure}

\begin{multicols}{2}

It is obvious from eq. (\ref{eq1d}) that by a proper choice
of the sample parameters one can tune
the effective temperature range at which saturation feature occurs on the
log-log plot. E.g. {\it thinner} samples with identical other parameters will
show no saturation down to {\it lower} temperatures. 
The lack of saturation for such samples even at relatively low $T$
would, of course, {\it not}
imply that the electron dephasing rate is zero at $T=0$. 

Let us replot the "non-saturating" data \cite{SM1,SM2} on a linear 
scale. In Fig. 3 we collected the data points for different samples studied
in different experiments ($Au1$, $Au2$, $Au6$ of Ref. \cite{Webb},
$Ag$ of Ref. \cite{SM1} and $AuMSU$ of Ref. \cite{SM2}). All these
samples show essentially {\it the same} behavior at low temperatures.
Both the slope  $d(1/\tau_{\varphi})/dT$ and the values $\tau_{\varphi 0}$ 
slightly differ from sample to sample, but these differences are
unimportant and can easily be accounted for by different values of 
$D$ and $tw$ as well as by measurement uncertainties. 
Much more importantly, at low temperatures  $1/\tau_{\varphi}$ depends 
{\it linearly} on $T$ and clearly extrapolates to a {\it nonzero} value 
for all the samples presented in Fig. 3. This is exactly what
one expects from our eq. (\ref{eq1d}). Also the similarity 
with the results \cite{Nat} (our Fig. 1) is quite obvious.

In summary, we demonstrated that the experimental results
\cite{Nat,SM1,SM2} are in a quantitative agreement with our
theory \cite{GZ98} which predicts the low temperature saturation
of the electron dephasing time in diffusive conductors. 
No material dependence of $\tau_{\varphi}$ needs to be assumed in
order to explain the data \cite{SM1,SM2} for the samples $Ag$, $Cu$
and $AuMSU$.

\end{multicols}


\begin{references}
\bibitem{FN} This cutoff procedure is correct for sufficiently
good metals which we only consider here. In systems like, e.g., 
granular or strongly disordered polycrystalline materials 
$\omega_c$ (as well as other details) can be different.
\bibitem{FN1} Eq. (\ref{eq1d}) gives correct but slightly
oversimplified dependence of $\tau_{\varphi}$ on $T$ in the
intermediate temperature regime where thermal and quantum
contributions are comparable. This inaccuracy (related
to the way of definition of $\tau_{\varphi}$) is, however, totally unimportant
for our conclusions.
\end{references}
\end{document}